\documentclass[aps,pre,twocolumn,reprint,superscriptaddress,citeautoscript]{revtex4-2}

\usepackage{graphicx}
\usepackage{dcolumn}
\usepackage{bm}

\usepackage{xcolor}
\usepackage{soul}
\usepackage{graphicx}
\usepackage{dcolumn}
\usepackage{bm}
\usepackage{mathrsfs}
\usepackage{amsmath}
\def\lesssim{\ \raise.3ex\hbox{$<$}\kern-0.8em\lower.7ex\hbox{$\sim$}\ }
\def\gesim{\ \raise.3ex\hbox{$>$}\kern-0.8em\lower.7ex\hbox{$\sim$}\ }

\begin{document}
\title{Optimal control approach to Olympic weightlifting exercise: \\Minimal model of the snatch pull}
\author{Hiroyuki Tajima}
\affiliation{Department of Physics, Graduate School of Science, The University of Tokyo, Tokyo 113-0033, Japan}
\affiliation{RIKEN Nishina Center, Wako 351-0198, Japan}
\affiliation{Quark Nuclear Science Institute, The University of Tokyo, Tokyo 113-0033, Japan}
\author{Hideyuki Nagao}
\affiliation{Sakushin-Gakuin University, Utsunomiya 321-3295, Japan}
\author{Kenya Tanaka}
\affiliation{Department of Medicine, Graduate School of Medicine, The University of Tokyo, Tokyo 113-0033, Japan}
\author{Hideaki Nishikawa}
\affiliation{Department of Physics, Kyoto University, Kyoto 606-8502, Japan}
\affiliation{Analytical Quantum Complexity RIKEN Hakubi Research Team, RIKEN Center for Quantum Computing (RQC), 2-1 Hirosawa, Wako, Saitama 351-0198, Japan}
\author{Eishiro Murakami}
\affiliation{Department of Sport Sciences,
Graduate School of Sport Sciences, Waseda University, Tokorozawa 359-1192, Japan}
\author{Akito Ida}
\affiliation{Department of Life Sciences, Graduate School of Arts and Sciences, The University of Tokyo, Tokyo 153-8902, Japan}
\author{Masataka Watanabe}
\affiliation{Department of Physics, Graduate School of Science, The University of Tokyo, Tokyo 113-0033, Japan}

\date{\today}
\begin{abstract}
  We theoretically investigate the biomechanical aspects of Olympic weightlifting within the framework of optimal control theory.
  The squared force and the rate of force development (RFD) defined by the time derivative of the force are taken into account in the evaluation functions of the first and second pull phases of the snatch motion.
  Focusing on the vertical trajectory of the barbell, we develop a minimal model to describe the mechanical characteristics of the weightlifting exercise. 
  The calculated barbell trajectory agrees well with the experimental data obtained by video analysis.
  Our study would be useful for the further development of mathematical models for weightlifting motions and related exercises.
\end{abstract}

\maketitle

\section{Introduction}
\begin{figure}[t]
    \centering
    \includegraphics[width=0.9\linewidth]{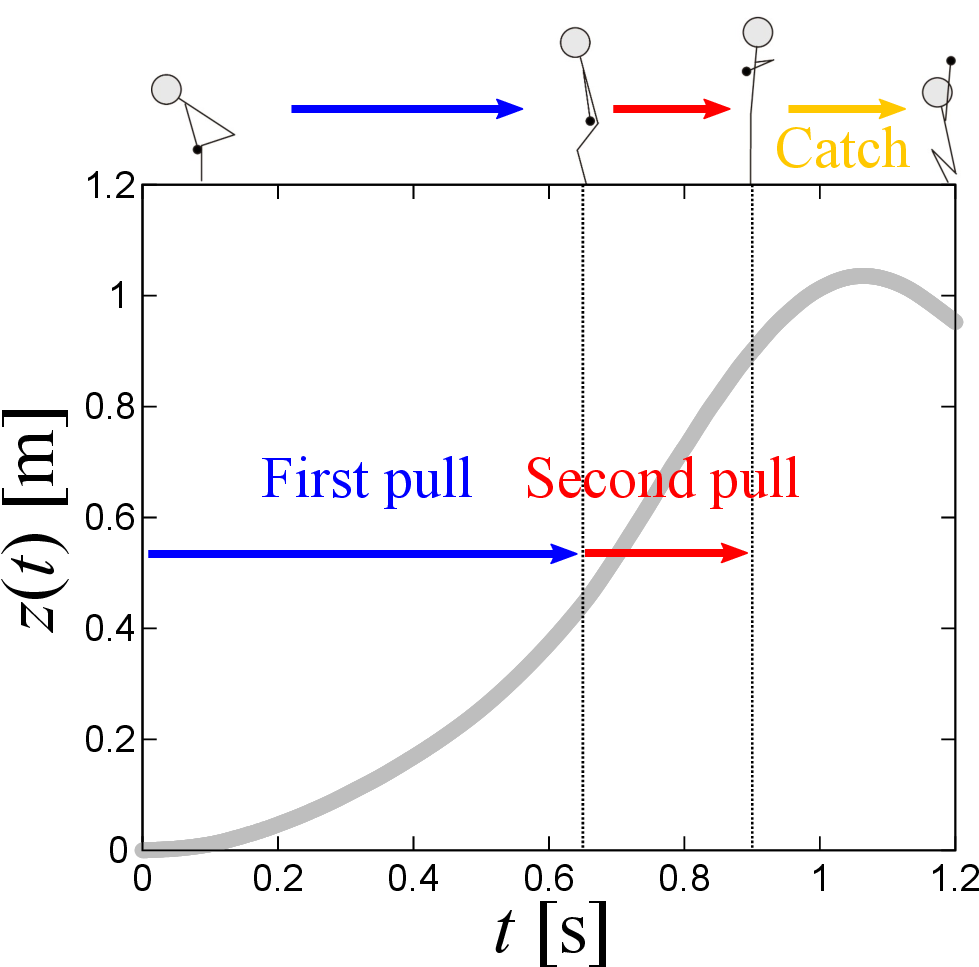}
    \caption{Observed vertical position of the bar during the snatch in Ref.~\cite{nagao2022validation}. At the beginning, an athlete pulls the bar relatively slowly to prepare for a strong drive at an appropriate posture. This phase is called ``first pull." In turn, at the appropriate posture which might be close to that of the vertical jump, an athlete use the whole muscle strength to induce a strong force on the bar, called ``second pull." After these pulling phases, an athlete tries to catch the bar. }
    \label{fig:1}
\end{figure}

Olympic weightlifting, consisting of snatch and clean \& jerk, is one of the most fundamental exercises for modern athletes.
 In recent decades, the biomechanics of Olympic weightlifting~\cite{Kauhanen1984,baumann1988snatch,bartonietz1996biomechanics} has attracted much attention to improve the performance not only of weightlifting itself but also of other sports conditioning~\cite {hedrick2008weightlifting}.
In particular, the pull motion plays a crucial role in both snatch and clean \& jerk~\cite{enoka1979pull} (see also details about snatch in Fig.~\ref{fig:1}), and is moreover useful for plyometric training of athletes.
Since the motion itself is apparently simple compared to other sports movements (but involves sophisticated techniques developed by intensive workouts), it can also be a good testing ground for the mechanical and mathematical studies.

Recently, the importance of the so-called ratio of force development (RFD) has been widely recognized in the communities of strength training~\cite{Cormie2007,Cormie2010a,Cormie2010b,Nuzzo2008,Baker2001a,Stone2003,Cormie2011,Baker2001b}.
The RFD may be defined as the time derivative of the force $F(t)$, namely,
\begin{align}
\label{eq:RFD}
    {\rm RFD}(t)=\frac{dF(t)}{dt},
\end{align}
which characterizes the explosive power of an athlete (note that RFD may also be defined as the change of $F(t)$ per unit time instead of the time derivative in the literature).
It is no doubt that the explosive power and thus RFD are important in Olympic weightlifting.
However, it has not been well explored how RFD can be mathematically taken into account in studies of the biomechanics of Olympic weightlifting.

To this end, the optimal control theory is a useful tool for investigating sports biomechanics in an efficient way. 
Optimal control is one of the central concepts in robotics~\cite{gasparetto2015path}, human motor systems~\cite{PhysRevE.83.031927}, as well as quantum physics~\cite{PhysRevA.68.062308,PhysRevLett.103.240501,PhysRevX.8.021059}.
In this connection, the analysis based on the one-dimensional Newtonian equation of motion has been used for optimizing a vertical jump of a robot~\cite{PhysRevLett.109.174301}.
Notably, the weightlifting pull motion and the vertical jump squat share a similarity in terms of the force-velocity relationship~\cite{sandau2023optimal,takei2024portions},
implying that the one-dimensional description may be useful even for the weightlifting motion although it is known that the pull in the weightlifting exhibits a specific bar trajectory in the sagittal plane~\cite{Kauhanen1984,baumann1988snatch,bartonietz1996biomechanics}.
From a physical viewpoint, as it is intuitive to understand  essences of the weightlifting motion as simple as possible, the one-dimensional mechanical analysis on the pull would be worth investigating.
It would also be beneficial for further developments of the weightlifting technique~\cite{nagao2019biomechanical,nagao2023biomechanical}.

In this paper, we examine the barbell trajectory during the snatch motion in terms of the optimal control approach.
In particular, we focus on the optimization of the force $F(t)$ as well as RFD given by $dF(t)/dt$, which can be reinterpreted as the path optimization via the Newtonian equation of motion without friction.
To consider the characteristics of the snatch motion consisting of first and second pull phases as shown in Fig.~\ref{fig:1},
we employ two evaluation functions responsible for the first and second pull phases in analogy with the squat model~\cite{matsui2016mathematical}.

This paper is organized as follows.
In Sec.~\ref{sec:formalism},
we present a general framework of a simplified mathematical model describing the vertical barbell trajectory during the snatch pull.
In Sec.~\ref{sec:results},
we compare our results of acceleration and jerk optimization models with the previous measurement of the barbell trajectory.
Finally, we give a summary of this paper and discuss future perspectives in Sec.~\ref{sec:summary}.

\section{Formalism}
\label{sec:formalism}

We consider the one-dimensional vertical motion of the barbell described by the Newtonian equation of motion
\begin{align}
\label{eq:1}
    m\frac{d^2z(t)}{dt^2}=F(t)-mg,
\end{align}
where $m$ and $z(t)$ are a mass and a vertical position of the bar, respectively.
$F(t)$ is an external force induced by an athlete and 
$g=9.80 \, {\rm m/s^2}$ is the gravitational acceleration constant.

The barbell trajectory is governed by $F(t)$.
In this regard, the problem is reduced to the optimal control of $F(t)$.
On the other hand, the optimization schemes can empirically be different in the first and second pull phases.
To this end, it is useful to introduce the evaluation function $J$ given by
\begin{align}
\label{eq:2}
    J=J_1+J_2,
\end{align}
where $J_{k=1}$ and $J_{k=2}$ are the evaluation functions in the first pull phase ($k=1$) and the second pull phase ($k=2$), respectively.
These can be expressed as
\begin{align}
\label{eq:3}
    J_{k}=
    \int_{t_{k,{\rm s}}}^{t_{k,{\rm f}}}
    dt\,
    \sum_{n=0}^{N_k}
    w_{k,n}
    \tau_k^{2n}
    \left[\frac{d^nF(t)}{dt^n}\right]^2,
\end{align}
where $\omega_{k,n}$ is a weight parameter that characterizes the relative importance of $n$-th derivative of $F(t)$ during the motion ($n=0,1,2,\cdots,N_k$, where $N_k$ is the cutoff number). 
In particular, $w_{k,0}$ and $w_{k,1}$ are the weight parameters for forces and RFD (defined in Eq.~\eqref{eq:RFD}), respectively.
In Eq.~\eqref{eq:3}, $t_{k,{\rm s}}$ and $t_{k,{\rm f}}$ are the initial and final times of each phase, respectively.
$\tau_k=t_{k,{\rm f}}-t_{k,{\rm s}}$ denotes the period of each phase.
Using Eq.~\eqref{eq:1}, one can rewrite
Eq.~\eqref{eq:3} as
\begin{align}
    J_{k}&=
    m^2w_{k,0}
    \int_{t_{k,{\rm s}}}^{t_{k,{\rm f}}}
    dt\,
    \left( g +z^{[2]}\right)^2\cr
    &
    \quad\quad
    +
    \int_{t_{k,{\rm s}}}^{t_{k,{\rm f}}}
    dt\,
    \sum_{n=1}^{N_k}
    m^2w_{k,n}\tau_k^{2n}
    \left(z^{[n+2]}\right)^2,
\end{align}
where we introduced $z^{[n]}=\frac{d^{n}z(t)}{dt^n}$ for convenience.
In this way, one can find that the optimization of $J$ with $F(t)$ is now equivalent to that with $z^{(n)}$ thanks to the simplification of the theoretical model.
For instance, the RFD optimization is rewritten as the jerk optimization. Note that this can be a simplified analogue of the equivalence between the minimum-jerk  and minimum-torque change models, which have also been discussed in the Riemannian geometry~\cite{PhysRevE.83.031927}.
Then, we consider the stationary condition $\delta J_k=0$, leading to the generalized Euler-Lagrange equation with higher derivatives of $z$ given by
\begin{align}
\label{eq:5}
    \sum_{n=0}^{N_k}
    (-1)^n\frac{d^n}{dt^n}
    \left(\frac{\partial\mathcal{L}_k}{\partial z^{[n]}}\right)=0,
\end{align}
where
\begin{align}
\label{eq:6}
    \mathcal{L}_k=m^2w_{k,0}\left(g+z^{[2]}\right)^2 +\sum_{n=1}^{N_k}m^2w_{k,n}\tau_k^{2n}\left(z^{[n+2]}\right)^2
\end{align}
is the effective Lagrangian in the first  ($k=1$) and second ($k=2$) pull phases.
From Eqs.~\eqref{eq:5} and \eqref{eq:6},
we obtain
\begin{align}
\label{eq:7}
    \sum_{n=2}^{N_k+2}2(-1)^n w_{k,n-2}\tau_k^{2(n-2)} z^{[2n]}=0.
\end{align}

Practically, we consider the optimization of the acceleration $z^{[2]}$ and the jerk $z^{[3]}$ in the following (i.e., $N_k=1$).
In such a case,
Eq.~\eqref{eq:7} is reduced to
\begin{align}
\label{eq:8}
    w_{k,0}\frac{d^4z(t)}{dt^4}-w_{k,1}\tau_k^2\frac{d^6z(t)}{dt^6}=0.
\end{align}
The solution of Eq.~\eqref{eq:8}
for $t_{k,{\rm s}}\leq t\leq t_{k,{\rm f}}$
is obtained analytically as
\begin{align}
\label{eq:9}
    z(t)&=\alpha_{k,0}+\alpha_{k,1}(t-t_{k,{\rm s}})\cr
    &\quad
    +\alpha_{k,2}(t-t_{k,{\rm s}})^2
    +\alpha_{k,3}(t-t_{k,{\rm s}})^3\cr
    &\quad
    +\beta_{k,+}e^{\lambda_k\frac{t-t_{k,{\rm s}}}{\tau_k}}
    +\beta_{k,-}e^{-\lambda_k\frac{t-t_{k,{\rm s}}}{\tau_k}},
\end{align}
where $\lambda_k=\sqrt{w_{k,0}/w_{k,1}}$ is the squared ratio of the weight parameters.
$\alpha_{k,0}$, $\alpha_{k,1}$, $\alpha_{k,2}$, $\alpha_{k,3}$, $\beta_{k,+}$, and $\beta_{k,-}$ are constants to be determined by the boundary conditions.

Furthermore, it is useful to consider the limits of $\lambda_{k}\rightarrow \infty$
and $\lambda_{k}\rightarrow 0$.
These correspond to the acceleration optimization and the jerk optimization, respectively.
For the case with $\lambda_{k}\rightarrow \infty$,
we need to set $\beta_{k,+}=0$ to obtain $z(t)$ without the divergence (i.e., $e^{\lambda_k\frac{t-t_{k,{\rm s}}}{\tau_k}}\rightarrow \infty$).
Meanwhile, $\beta_{k,-}$ becomes irrelevant due to $e^{-\lambda_k\frac{t-t_{k,{\rm s}}}{\tau_k}}\rightarrow 0$
Eventually, we obtain
\begin{align}
\label{eq:10}
    z(t)&\simeq \alpha_{k,0}+\alpha_{k,1}
    (t-t_{k,{\rm s}})
    +\alpha_{k,2}(t-t_{k,{\rm s}})^2\cr
    &
    \quad 
    +\alpha_{k,3}(t-t_{k,{\rm s}})^3.
\end{align}
On the other hand, for the case with $\lambda_{k}\rightarrow 0$,
substituting Eq.~\eqref{eq:9} into Eq.~\eqref{eq:8},
we find
\begin{align}
\lim_{\lambda_k\rightarrow 0}
    (\beta_{k,+}+\beta_{k,-})\left(\frac{\lambda_k}{\tau_k}\right)^6=0,
\end{align}
indicating that $\beta_{k,\pm}=O(\lambda_k^{-5})$.
Expanding $e^{\pm\lambda_k\frac{t-t_{k,{\rm s}}}{\tau_k}}$ up to order $\lambda_k^5$ and rearranging the constants, we obtain the analytical expression of the jerk optimization model given by
\begin{align}
\label{eq:12}
    z(t)\simeq &
    \,\alpha_{k,0}+\alpha_{k,1}(t-t_{k,{\rm s}})
    +\alpha_{k,2}(t-t_{k,{\rm s}})^2\cr
    &
    \,
    +\alpha_{k,3}(t-t_{k,{\rm s}})^3
    +\alpha_{k,4}(t-t_{k,{\rm s}})^4\cr
    &\,+\alpha_{k,5}(t-t_{k,{\rm s}})^5,
\end{align}
where $a_{k,1\cdots5}$ are determined by the boundary conditions.

\begin{table}[bt]
\begin{center}
\caption{Boundary conditions for the first and second pull phases in the snatch motion.}
\label{tab:bc}
  \begin{ruledtabular}
\begin{tabular}{c|c|c|c}
    $t$ & ${z}(t)$ & $\dot{z}(t)$ &$\ddot{z}(t)$ \\
    \hline 
    $t=t_{1,{\rm s}}=0$ & $0$ & $0$ & $a_0$ \\
    $t=t_{1,{\rm f}}=t_{2,{\rm s}}=T_1$ & $h_1$ & $v_1$ & $a_{1}$ \\
    $t=t_{2,{\rm f}}=T_1+T_2$ & $h_2$ & $v_{2}$ & $a_2$
\end{tabular}
\end{ruledtabular}
\end{center}
\end{table}

In Table~\ref{tab:bc}, we summarize the boundary conditions for determining the parameters in each optimization scheme.
At the initial time $t=t_{1,{\rm s}}$ of the first pull (taken to be $t_{1,{\rm s}}=0$ for convenience),
we assume $z(t_{1,{\rm s}})=\dot{z}(t_{1,{\rm s}})=0$, leading to
\begin{align}
    \alpha_{1,0}=\alpha_{1,1}=0.
\end{align}
regardless of the optimization schemes (i.e., acceleration and jerk minimizations given by Eqs.~\eqref{eq:10} and \eqref{eq:12}).
Meanwhile, the experimental data~\cite{nagao2022validation} suggests a nonzero acceleration at $t=0$.
In this regard, we keep
\begin{align}
    \alpha_{1,2}=a_0\neq0,
\end{align}
where $a_0$ is an initial acceleration.
At $t=t_{1,{\rm f}}$, the phase changes from the first pull to the second one with a given height $h_1$, velocity $v_1$, and acceleration $a_1$. In this sense, one can find $t_{2,{\rm s}}=t_{1,{\rm f}}=T_1$ where $T_1$ is the period of the first pull.
Eventually, the second pull phase lasts up to $t=t_{2,{\rm f}}=T_1+T_2$ ($T_2$ is the period of the second pull) with $h_2$, $v_2$, and $a_2$. After that, the barbell motion follows the free fall because the external force is no longer applied during the catch phase.
Note that the free-fall approximation works well empirically in the catch phase after the snatch pull~\cite{sandau2021predictive}. 

In the following, we show the analytical expression of the vertical barbell trajectory within acceleration and jerk optimization schemes in the first and second pull phases.

\subsection{Acceleration optimization in the first pull}
For the acceleration optimization scheme in the first pull phase,
the two parameters $a_{1,2}$ and $a_{1,3}$ are determined by the boundary conditions. 
To this end, we employ $a_0$ and $h_1$, which are relatively easy to deduce from the experimental data compared to $v_1$ and $a_1$.
We obtain
\begin{align}
\label{eq:15}
    z_{\rm A}(t)=
    \frac{1}{2}a_0 t^2+
    \left(h_1-\frac{1}{2}a_0T_1^2\right)\left(\frac{t}{T_1}\right)^3. 
\end{align}
Simultaneously, we can obtain $v_1$ from Eq.~\eqref{eq:15} as
\begin{align}
\label{eq:16}
    v_1 =\frac{1}{T_1}\left(3h_1-\frac{1}{2}a_0T_1^2\right).
\end{align}
We note that $v_1$ in Eq.~\eqref{eq:16} does not necessarily coincide with the experimental data.
However, 
for the sake of the comparison between the acceleration and jerk optimization schemes,
it is useful to employ $v_1$ as the boundary condition for the other schemes.

\subsection{Jerk optimization in the first pull}
In the present case, the general form of the trajectory is given by
\begin{align}
    z_{\rm J}(t)=
    \frac{1}{2}a_0 t^2+
    \alpha_{1,3}t^3
    +\alpha_{1,4}t^4
    +\alpha_{1,5}t^5,
\end{align}
where $\alpha_{1,3}$, $\alpha_{1,4}$, and $\alpha_{1,5}$ are determined by the boundary conditions
\begin{align}
&    z_{\rm J}(t=T_1)=h_1,
\quad
    \left.\frac{d{z}_{\rm J}(t)}{dt}\right|_{t=T_1}=v_1,\cr
&\quad
    \left.\frac{d^2{z}_{\rm J}(t)}{dt^2}\right|_{t=T_1}=a_1.
\end{align}
After the straightforward calculations,
we obtain
\begin{widetext}
    \begin{align}
    z_{\rm J}(t)
    &=\frac{a_0t^2}{2}+\left[
    10h_1
   -4v_1T_1
   +\frac{1}{2}a_1T_1^2-\frac{3}{2}a_0T_1^2
    \right]
    \left(\frac{t}{T_1}\right)^3
    +\left[7v_1T_1 -15h_1
   -a_1T_1^2
   +\frac{3}{2}a_0T_1^2
   \right]
   \left(\frac{t}{T_1}\right)^4\cr
   &\quad +
   \left[
   \frac{1}{2}(a_1-a_0)T_1^2
   -3v_1 T_1+6h_1
   \right]
   \left(\frac{t}{T_1}\right)^5.
\end{align}
\end{widetext}

\subsection{Acceleration optimization in the second pull}
We have four coefficients to be determined by the boundary conditions as
\begin{align}
    z_{\rm A}(t)&=\alpha_{2,0}+\alpha_{2,1}(t-t_{2,{\rm s}})
    +\alpha_{2,2}(t-t_{2,{\rm s}})^2
    \cr
    &\quad
    +\alpha_{2,3}(t-t_{2,{\rm s}})^3.
\end{align}
Here
we employ $h_1$, $h_2$, $v_1$, and $v_2$ for the boundary conditions.
Note that these parameters are easy to deduce from the experimental data compared to $a_1$ and $a_2$, which require further numerical time derivative of the observed velocity.
In such a case, we obtain
\begin{align}
    z_{\rm A}(t)
    &=h_1 + v_1 (t-t_{2,{\rm s}})
    \cr
    &\quad+[3(h_2-h_1)-(v_2+2v_1)T_2]\left(\frac{t-t_{2,{\rm s}}}{T_2}\right)^2\cr
    &\quad
    +[(v_2+v_1)T_2-2(h_2-h_1)]\left(\frac{t-t_{2,{\rm s}}}{T_2}\right)^3.
\end{align}

\subsection{Jerk optimization in the second pull }
Finally, we obtain an analytical formula for the jerk optimization in the second pull phase, which require six boundary conditions to determine all the parameters.
Accordingly, we use $h_1$, $h_2$, $v_1$, $v_2$, $a_1$, and $a_2$ for the boundary conditions.
Eventually, we obtain
\begin{widetext}
\begin{align}
    z_{\rm J}(t)=&h_1 
    +v_1(t-t_{2,{\rm s}})
    +\frac{1}{2}a_1 (t-t_{2,{\rm s}})^2
    +\left[10(h_2-h_1)-(4v_2+6v_1)T_2
    +\frac{\left(a_2-3a_1\right)T_2^2}{2}\right]\left(\frac{t-t_{2,{\rm s}}}{T_2}\right)^3\cr
    &+
    \left[
    -15(h_2-h_1)
    +(7v_2+8v_1)T_2
    -\left(a_2-\frac{3}{2}a_1\right)T_2^2
    \right]\left(\frac{t-t_{2,{\rm s}}}{T_2}\right)^4\cr
    &+\left[
    6(h_2-h_1) -3(v_2+v_1)T_2+\left(\frac{a_2-a_1}{2}\right)T_2^2
    \right]\left(\frac{t-t_{2,{\rm s}}}{T_2}\right)^5.
\end{align}
\end{widetext}


\section{Results and discussion}
\label{sec:results}
Combining the acceleration and jerk optimization models developed in Sec.~\ref{sec:formalism}, we discuss the vertical barbell trajectory during the snatch motion.
In Table~\ref{tab:bc2}, we summarize the parameters estimated from the observed barbell trajectory in Ref.~\cite{nagao2022validation}.
\begin{table}[bt]
\begin{center}
\caption{Parameters estimated from Ref.~\cite{nagao2022validation}. The third and fourth columns indicate the model schemes (A: acceleration optimization, J: jerk optimization) which employ each parameter.}
\label{tab:bc2}
  \begin{ruledtabular}
\begin{tabular}{cccc}
Parameter & Numerical value & First pull & Second pull \\
\hline
$T_1$& $0.65 \, {\rm s}$  & A, J & -
\\

$a_0$ & $2.10 \, {\rm m/s^2}$ & A, J & - \\ 
$h_1$ & $0.44 \, {\rm m}$ & A, J & A, J \\
$v_1$ & $\frac{1}{T_1}\left(3h_1-\frac{a_0T_1^2}{2}\right)$ & A, J & A, J \\ 
$a_1$ & $0 $ & J & J \\ 
$T_2$& $0.25 \, {\rm s}$ & - & A, J
\\
$h_2$ & $0.90 \, {\rm m}$ & - & A, J \\ 
$v_2$ & $1.45 \, {\rm m/s}$ & - & A, J \\ 
$a_2$ & $-g $ & - & J \\ 
\end{tabular}
\end{ruledtabular}
\end{center}
\end{table}
Since it is difficult to obtain the accelerations $a_1$ and $a_2$ from the motion capture data~\cite{nagao2022validation},
in this work we assume that $a_1\simeq 0$ and $a_2\simeq -g$.
The former assumption is based on the fact that the barbell acceleration is quenched at the transition point between the first and second pulls ($t=t_{1,{\rm f}}=T_1$), because weightlifters must adjust their posture to avoid contact between the bar and the knees at this moment.
In this work, we estimate $T_1$ from the horizontal velocity of the barbell in Ref.~\cite{nagao2022validation}, which shows a clear signature of this transition as the rapid change of the horizontal jump.
Note that such a tendency can also be found in the ground \-reaction force via the force-plate measurements (e.g., see Ref.~\cite{chavda2025effect}). 
The latter assumption is associated with the fact that the bar drive ends at $t=t_{2,{\rm f}}=T_1+T_2$ and instead weightlifters start dropping under the bar to catch the bar.
In this catch phase, the vertical barbell motion is approximately described by the free fall with $F(t)=0$, that is,
\begin{align}
    z(t>t_{2,{\rm f}})\simeq h_2+v_2(t-t_{2,{\rm f}})+\frac{1}{2}g(t-t_{2,{\rm f}})^2.
\end{align}
In this work, $T_2$ is estimated based on the linear fit of the vertical barbell velocity around $t= 1\,{\rm s}$.

\begin{figure}[t]
    \centering
    \includegraphics[width=0.8\linewidth]{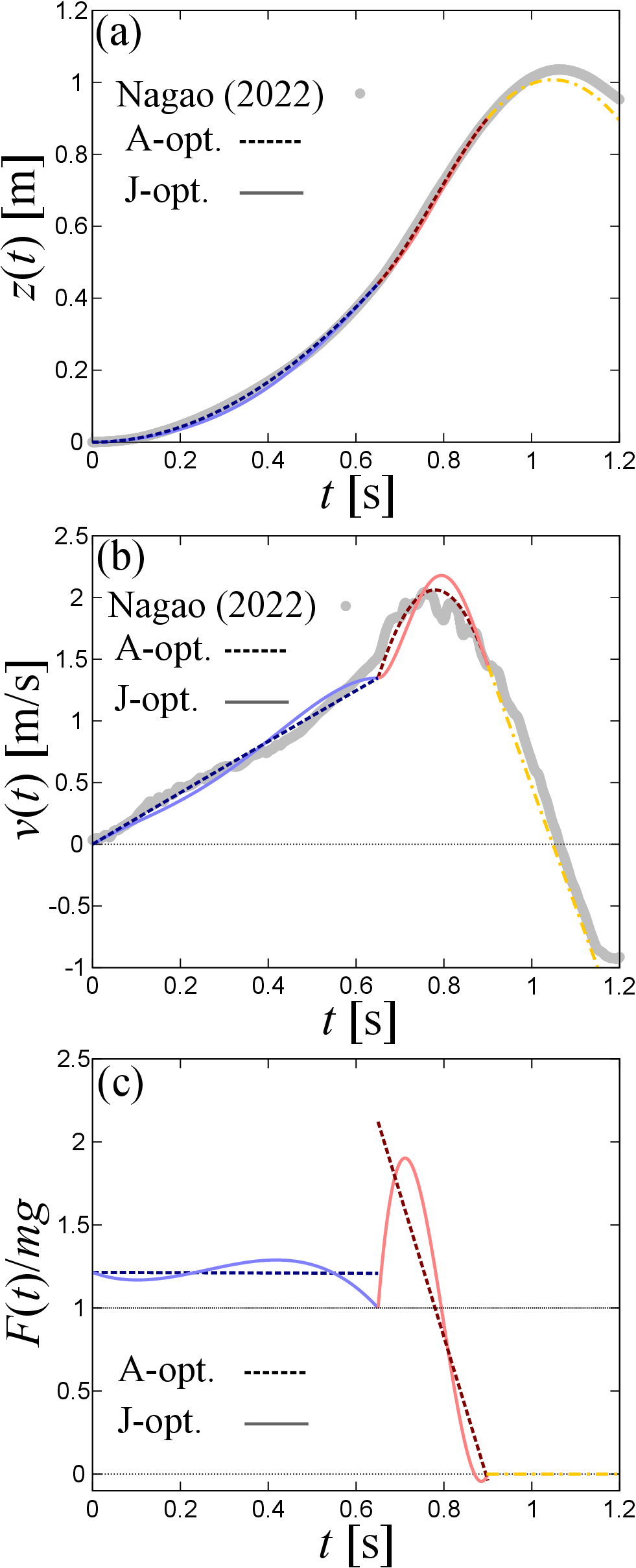}
    \caption{(a) Vertical position $z(t)$, (b) velocity $v(t)$, and (c) force $F(t)$ during the snatch motion.
    The gray circles show the experimental data in Ref.~\cite{nagao2022validation}.
    The solid and dashed curves represent the acceleration optimization model (A-opt.) and the jerk optimization model (J-opt.), respectively.
    The chain-dot curve at $t\geq t_{2,{\rm f}}=0.90\,{\rm s}$ corresponds to the free fall motion with $F(t)=0$ in the catch phase.
    }
    \label{fig:2}
\end{figure}

Figures~\ref{fig:2}(a) and (b) show the barbell trajectory $z(t)$ and the velocity $v(t)$ in the snatch motion obtained by the acceleration and jerk optimization models.
For comparison, we also present the observed data in Ref.~\cite{nagao2022validation}.
Note that, in Fig.~\ref{fig:2}(c), only the theoretical results of the force $F(t)=m\left[g+\frac{d^2z(t)}{dt^2}\right]$ acting on the barbell are shown because it is difficult to obtain the acceleration from the motion capture data by numerical differentiation.
In particular, the small oscillation of $v(t)$ in the experimental data is associated with the elastic vibration of the bar, which is one of the unavoidable sources of noise in the acceleration.

In general, the experimental data of $z(t)$ in Fig.~\ref{fig:2}(a) are well explained by both models,
regardless of the first and second pull phases. A small deviation of $z(t)$ in the catch phase ($t\geq t_{2,{\rm f}}=0.9\,{\rm s}$) may be related to the numerical errors in the estimation of $T_2$ and $v_2$ due to the small oscillating behavior of the observed $v(t)$.
However, a more accurate description of this phase is beyond the scope of this paper.

The difference between the acceleration and jerk optimization models is clearer in $v(t)$ shown in Figs.~\ref{fig:2}(b).
One can find that the acceleration optimization scheme exhibits a linear increase in the first pull phase. 
While the jerk optimization scheme also shows a linear behavior in the first pull phase, the velocity is slightly slower near the initial time and faster near the transition point, compared to the acceleration optimization case.
In the second pull phase, while both optimization schemes show an upper convex behavior, the maximum velocity in the jerk optimization scheme is larger than the acceleration one.
A rapid increase of the observed $v(t)$ near the transition point can also be found in both schemes.
On the other hand, the experimental data is relatively close to the acceleration optimization scheme, suggesting that the optimization of $|F(t)|^2$ may play a more important role than that of $|dF(t)/dt|^2$ in the second pull phase.
Although still the difference between two schemes in the velocities is small and depends on the choice of the boundary conditions,
it is suggestive to see an interplay of impulse and RFD, which are somewhat related to $|F(t)|^2$ and $|dF(t)/dt|^2$, in the first and second pull phases.
Indeed, one can see the case of combining two optimization schemes in the different pull phases (e.g., acceleration and jerk optimizations in the first and second pull phases, respectively) by comparing two lines in each phase.
This is another advantage of dividing the evaluation functions into two phases in Eq.~\eqref{eq:2}.

Finally, one can see a clear difference of $F(t)$ between the acceleration and jerk optimization schemes as shown in Fig.~\ref{fig:2}(c). 
The acceleration optimization scheme exhibits a linear behavior of $F(t)$ given by
\begin{align}
    F(t)=
m\left[g+a_0+\frac{6t}{T_1^3}
    \left(h_1-\frac{1}{2}a_0T_1^2\right)
    \right], 
\end{align}
for $0\leq t\leq t_{1,{\rm f}}$, and
\begin{align}
    F(t)&=
m\biggl[
g+6(h_2-h_1)-2(v_2+2v_1)T_2\cr
  &\,\,\,\, +6[(v_2+v_1)T_2-2(h_2-h_1)]\left(\frac{t-t_{2,{\rm s}}}{T_2}\right)
\biggr],
\end{align}
for $t_{2,{\rm s}}\leq t\leq t_{2,{\rm f}}$.
Since the acceleration optimization scheme does not employ the boundary conditions of $a_1=0$ and $a_2=-g$,
$F(t)$ becomes discrete at $t=t_{1,{\rm f}}$ and $t=t_{2,{\rm f}}$, in contrast to the result of the jerk optimization scheme.
In the first pull phase, while the acceleration optimization scheme shows almost a time-independent force,
the jerk optimization scheme shows a small peak of $F(t)$.
Then, at the transition point, the latter scheme touches $F(t)=mg$ due to the boundary condition, and shows a rapid increase in the second pull phase.
Such a behavior can also be found in the typical measurement of the ground reaction force~\cite{chavda2025effect}.
However, it should be noted that the ground reaction force consists of not only the force applied to the barbell but also the force to rearrange the athlete's body.
In this sense, a detailed comparison between $F(t)$ and the ground reaction force will be left for future work.

Although the acceleration optimization scheme shows a jump at the transition point from the first pull phase to the second one, the global tendency of $F(t)$ is similar to the jerk optimization scheme.
Eventually, at $t=t_{2,{\rm f}}$, both schemes shows $F(t)\simeq 0$.
While the jerk optimization scheme gives exactly $F(t=t_{2,{\rm f}})=0$ due to the boundary condition, the acceleration optimization scheme also shows $F(t=t_{2,{\rm f}})\simeq 0$ even without the boundary condition.
We note that $F(t)$ becomes slightly negative near $t=t_{2,{\rm f}}$ in the jerk optimization scheme, implying that the boundary conditions obtained from the observed data is not ``optimal" in terms of the jerk optimization. Indeed, this negative region depends on the boundary conditions and thus we regard that it is not necessary to be taken into account seriously for our analysis.

\section{Summary and Future Perspective}
\label{sec:summary}
In summary,
we have discussed the barbell trajectory in the snatch exercise of Olympic weightlifting within the optimal control approach.
Using the evaluation function for the external force induced by a weightlifter (which is practically converted to optimization of the barbell trajectory via the equation of motion), we have examined how the observed barbell trajectory can be described by the acceleration and jerk optimization schemes.
It is found that both optimization schemes show good agreement with the experimental data, while their differences can be found in the time dependence of the force.
Our result indicates that weightlifters might try to achieve an ``optimized" form to lift the barbell in a given body environment.

For future perspectives,
while we focus on the one-dimensional motion of the barbell,
it is worth investigating the role of parameters in the evaluation functions in more realistic models including athlete's rigid bodies~\cite{nejadian2008optimization} as well as physiological properties such as force-length relation of muscles~\cite{gordon1966variation}.
Moreover, applications to other motions, such as deadlift and squat in powerlifting~\cite{radenkovic2018physics,radenkovic2020physics}, would be an interesting future direction.
Although we consider the currently available data for the barbell trajectory~\cite{nagao2022validation},
it is also important to examine different experimental data using the present approach.

\acknowledgements
H.~T. thanks Takahiro M. Doi, Sekiya Koike, Tetsuo Hatsuda, and Atsushi Nakamura for the useful discussion at the initial stage of this study. 
H.~T. is supported by JSPS KAKENHI for Grants (Nos.~JP22K13981, JP23K22429) from MEXT, Japan.
M.~W. is supported by a Grant-in-Aid for JSPS Fellows No.~22KJ1777, a Grant-in-Aid for Early-Career Scientists No.~25K17387, and by a MEXT KAKENHI Grant No.~24H00957.

\bibliographystyle{apsrev4-2}
\bibliography{reference.bib}

\end{document}